# Realization of Single Attosecond Pulse Generation in Multi-Cycle Regime through Modulation of Carrier Wave

*Xiang Yang[1], Yueping Niu[2,*], and Shangqing Gong[2,†]*

[1] *School of Computer Science and Technology, Henan Polytechnic University, Jiaozuo 454000, China*
[2] *Department of Physics, East China University of Science and Technology, Shanghai 200237, China*
Email: [*] *niuyp@ecust.edu.cn*, [†] *sqgong@ecust.edu.cn*

**Abstract:** An intuitive physical mechanism for single attosecond pulse generation in multi-cycle pulse regime is proposed. We suggest that the continuum spectra of HHG can be controlled through the modulation of carrier wave, while keeping the multi-cycle envelope unchanged approximately. By increasing the time interval between the adjacent maximum return kinetic energy peaks, the kinetic energy of the side peaks will be apparently suppressed by the multi-cycle envelope. The photons with energies near the cutoff burst only at the pulse center with high coherence, which contribute to the generation of single attosecond pulse. We verify this mechanism by using a multi-cycle tri-color pulse as driver field.

*OCIS codes*: 320.7110, 020.2649, 190.4160.

Recently, the generation of single attosecond (as) pulse has attracted great interests around the world, due to its wide applications in the detection and control of ultra-fast dynamic processes, such as Auge-decay process and real-time observation of the motion of an electronic wave packet in an atom [1]. Till now, attosecond pulses generation mainly come from the high-order harmonic generation (HHG) spectrum of atoms or molecules driven by a strong laser pulse [2, 3]. Usually, the HHG spectrum from a multi-cycle pulse contains only discrete odd harmonics, from which attosecond pulse train can be obtained [4]. When using a few-cycle laser pulse as driver field, the spectrum near the cutoff becomes continuum and then a single as pulse can be generated [5]. By this means, Goulielmakis *et al.* obtained a single 80as pulse which is the shortest pulse obtained in experiment at present [6]. However, the driver pulse they used was 3.3 fs with stable carrier-wave-envelop phase (CEP), which is still a challenge in recent experiment technique for many laboratories around the world. As an alternative way, many schemes have been proposed to generate single as pulse in multi-cycle regime. Corkum *et al.* proposed that using a driver pulse with polarization sweeping from circular through linear back to circular, i.e., the polarization gating control, a single as pulse can be obtained [7]. Lan *et al.* suggested that a single as pulse could be generated if using asymmetric target molecules [8]. Recently, we demonstrated that nonlinear chirped pulse can support isolate as pulse generation with duration almost independent of the chirp form and duration of the driver pulse [9].

In this letter, we provide an intuitive physical mechanism for the single attosecond pulse generation in the multi-cycle regime. According to the well-known three-step model (TSM) [10], when the electron collides with the parent ion, a photon is emitted with energy equal to the return kinetic energy (RKE) of the electron plus the ionization potential of the atom. For multi-cycle driver pulse, this process occurs every half-cycle and hence as pulse chain can be generated. While for few-cycle driver pulse, the side peaks of RKE are suppressed by the pulse envelop, and hence continuum spectrum appear and single as pulse can be generated. Hence, we think that in the multi-cycle regime if we keep the envelop unchanged approximately but modulate the carrier wave to increase the time interval between the adjacent maximum RKE peaks, the HHG cutoff can be controlled to be continuum and hence is suitable for single as pulse generation.

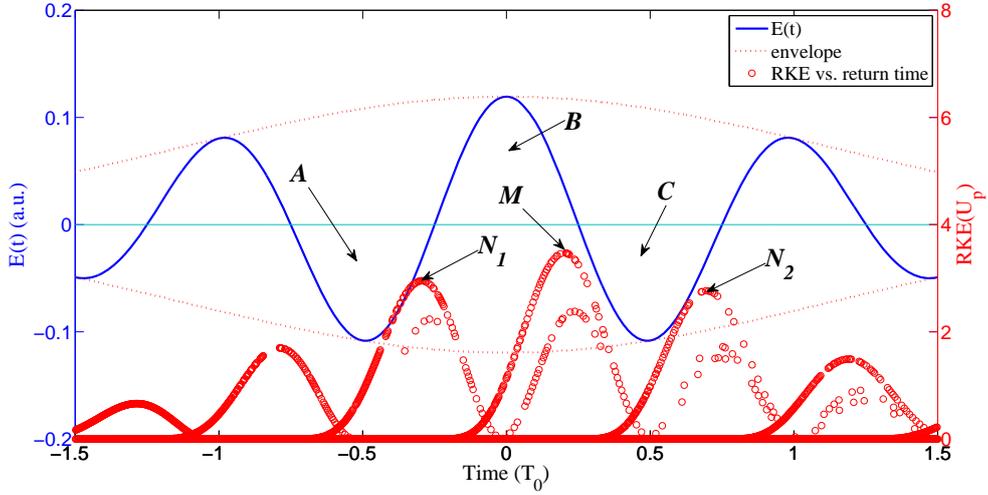

Fig.1 (Color online) The few-cycle laser pulse field with $\tau = 5 fs$ and the RKE map as a function of return time.

In order to illustrate our basic idea, we first review the classical trajectory of the electron in the laser field with a constant envelope. Just as described above, the RKE map repeats itself in each half cycle with a maximum $3.17 U_p$ ($U_p = E_0^2 / 4\omega_0^2$ is the ponderomotive in the laser field of strength $E_0$ and frequency $\omega_0$). This means that there are many burst times for each photon in the HHG spectrum. Therefore, the HHG spectrum is discrete from which only as pulse train can be obtained. In order to get single attosecond pulse, a continuum spectrum should appear in the HHG spectrum, that is to say, one should control the trajectory of the electron so that some photons have few burst times [11]. For linear polarized laser field, this can be realized by using a few-cycle pulse, such as $E(t) = E_0 f(t) \cos(\omega_0 t + \varphi_0)$, with $f(t) = e^{-2\ln 2 t^2 / \tau^2}$, $E_0 = 0.12 a.u.$ ($5 \times 10^{14}$ W/cm$^2$), $\omega_0 = 0.057 a.u.$ (800nm), $\tau = 5 fs$ and $\varphi_0 = 0$. Here, the atomic units (a.u.) are used. The classical results are shown in Fig. 1. It is clear that the impulse of the half cycle *A* and *C* is apparently decreased due to the influence of the envelope. Thereby, the maximum RKE of the electron which returns in the half cycle *A* (peak $N_1$) or *C* (peak $N_2$) is lower than that of the electron returns in the half cycle *B* (peak *M*), according to our analysis in Ref. [12]. The photons generated from the electrons with RKE between peak *M* and $N_1$ burst only in about $0.16T_0$ ($T_0 = 2\pi/\omega_0$) and the harmonics consist of these photons become continuous, from which a single as pulse can be generated.

It should be noted that the time interval between the above adjacent RKE peaks is $0.5T_0$, only the envelope with duration of few optical cycles can suppress the side RKE peaks, say $N_1$, $N_2$. With the increase of the pulse duration, the difference between peak *M* and $N_1$ ($N_2$) decreases smoothly. In order to obtain a relative greater difference of the RKE peaks in multi-cycle regime, one feasible way is to increase the time interval between the adjacent maximum RKE peaks. For a much clearer explanation, we first take a mathematically electric field

$$E_p(t) = \begin{cases} 2E_0 \cos(2\omega_0 t) & -T_0/8 \leq t \leq 3T_0/8, \\ 0 & -T_0/2 < t < -T_0/8 \text{ or } 3T_0/8 < t < T_0/2, \\ E_p(t - kT_0) & \text{otherwise}, k = \pm 1, \pm 2, \pm 3, \cdots. \end{cases} \quad (1)$$

as an example. Figure 2(a) shows $E_p(t)$ and its RKE vs. tunneling and return time. On one hand, we find that now the two RKE peaks return in the first and second half cycle (we mark them as *M* and *N*, the other peaks are not marked since the RKE map is periodic) have different values. The peak value of *M* is $5.77U_p$, which comes from the electron tunnels at time $t_{M'}$ (the time corresponding to the peak *M'*, the same below) and returns at $t_M$ in the first half cycle *A*. The peak value of *N* is $3.17U_p$, which comes from the electron tunnels at time $t_{N'}$ and returns at $t_N$ in the second half cycle *B*. Detailed investigation on the electric field shows that the electrons tunneling at $t_{M'}$ accelerates in the laser field and then moves with uniform velocity, which makes it move a longer distance from the parent ion than that tunnels at $t_{N'}$ before decelerates and returns to the parent ion. Thus, the electrons tunnel at $t_{M'}$ own greater RKE which makes the two RKE peaks *M* and *N* have different value. On the other hand, now the repeat period for the carrier wave is $T_0$. As a result, the time interval between the adjacent maximum RKE peaks is increased to $T_0$. Then, under the action of the multi-cycle envelop $f(t)$, the side peak $P_1$($P_2$) becomes much lower than the center peak *M* even in the multi-cycle regime, as shown in Fig. 2(b).

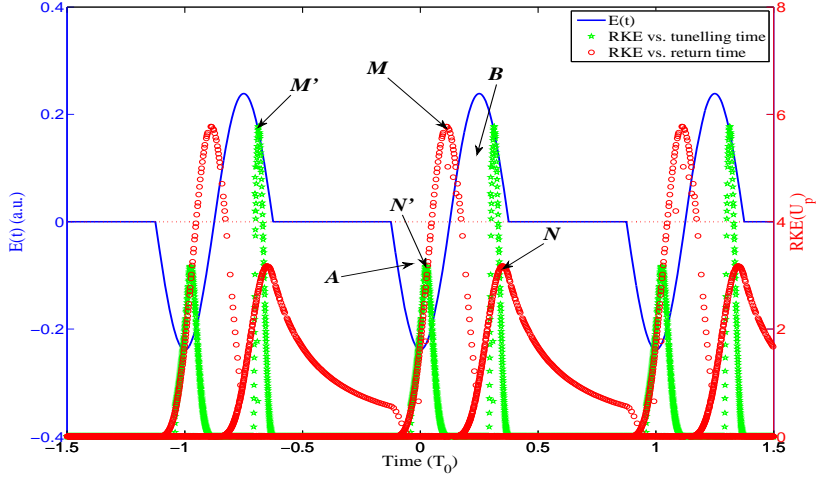

(a)

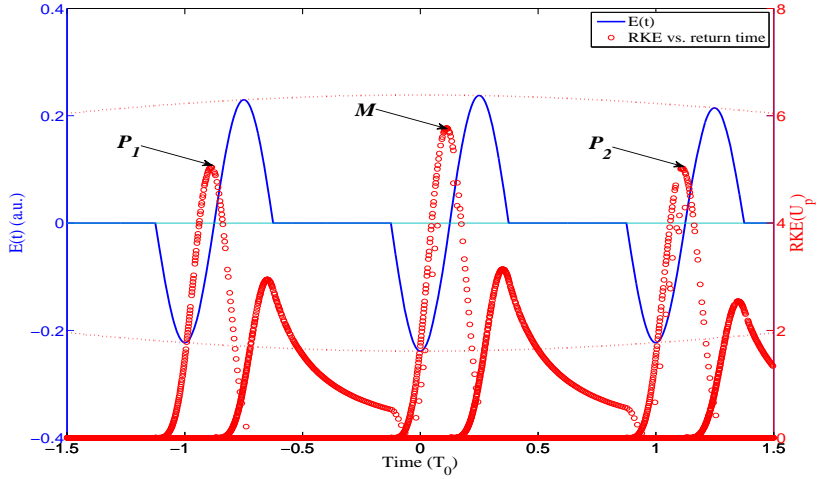

(b)

Fig. 2 (Color online) (a) The electric field $E_p(t)$ and the RKE vs. tunneling and return time. (b)The electric field $E_p(t)$ under the modulation of $f(t)$ with $\tau = 12\,fs$ and the RKE vs. return time.

In experiment, the electric field shown in Fig. 2(b) can be implemented approximately by the following tri-color laser pulse:

$$E_s(t) = E_0 f(t)\left[\sin(\omega_0 t + \varphi_0) + \sin(2\omega_0 t + 2\varphi_0) + \frac{1}{3}\sin(3\omega_0 t + 3\varphi_0)\right], \quad (2)$$

where $\varphi_0$ is set to be $-0.25\pi$. The electric field $E_s(t)$ is shown in Fig.3 (a). As is displayed in Fig. 3(b), the map of RKE is similar to that of Fig. 2(b). The maximum RKE (center peak $M$) is enhanced to $6.78U_p$ and the time interval becomes $T_0$. The HHG spectrum generated from the laser field $E_s(t)$ is shown in Fig.3 (c) (In our calculations, three-dimensional time-dependent Schrödinger equation (TDSE) with single active electron and dipole approximations is applied to describe the interaction between the helium atom and the laser field [13-15].). Its cutoff is at about 147th-order harmonic, which is consistent with the classical result shown in Fig.3 (b). Further, the electrons with RKE between center peak *M* and side peak *Q* return only in about $0.07T_0$, which means that the photons generated by these electrons burst with high coherence. By selecting the spectrum above the 129th-order harmonic with a high-pass filter, an isolated 117 attosecond pulse is obtained without phase compensation, as shown in Fig.3 (d).

Note that the repeat frequency of the tri-color field in Eq. (2) is $\omega_0$, so the time interval between peak *M* and *Q* in Fig. 3(b) is $T_0$. If we use a tri-color laser field with a lower repeat frequency, the time interval between peak *M* and *Q* should increase, which is more beneficial to the generation of single as pulse with longer pulse duration. Following this idea, another tri-color laser field with $\omega_0$, $2.2\omega_0$ and $3\omega_0$ is applied for test. Now the repeat frequency turns to be $\omega_0/5$ and the time interval between the adjacent maximum peaks is $2.5T_0$. Our numerical simulation shows that the side peaks of the RKE can be remarkably suppressed except that at the pulse center even with a 30fs multi-cycle pulse. By selecting the continuum spectrum above the 128th-harmonic, an isolated 116 attosecond pulse is obtained.

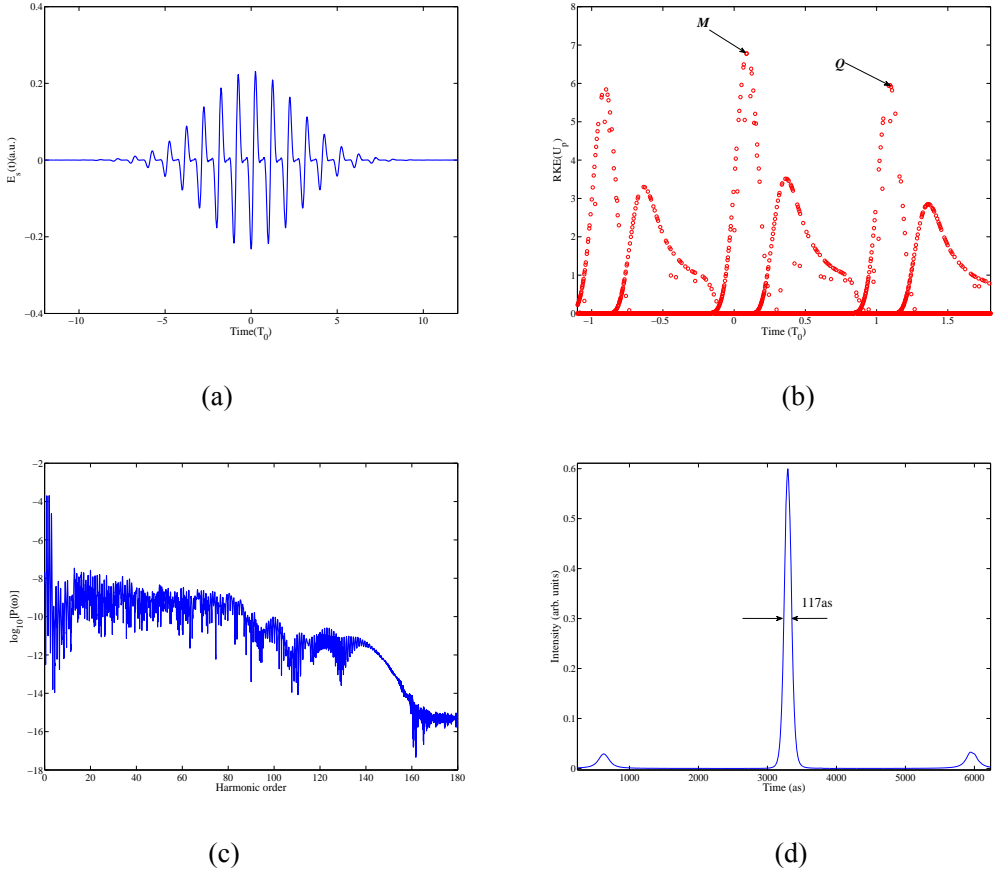

Fig. 3 (Color online) (a) The electric field $E_s(t)$ with $\tau = 12\,fs$. (b) The RKE of the electron driven by $E_s(t)$. (c) The HHG spectrum driven by the laser field in (a). (d) The generated

attosecond pulse.

In summary, an intuitive mechanism for the generation of single attosecond pulse has been proposed. By enlarging the time interval between the adjacent maximum RKE peaks, the kinetic energy of the side peaks are evidently suppressed by the pulse envelope even in the multi-cycle regime. The laser field can be implemented in experiment using a multi-cycle tri-color pulse. By adjusting the repeat frequency of the tri-color laser field, the time interval can be controlled and single attosecond pulse can be generated in multi-cycle regime with different pulse durations.

This work is supported by the National Basic Research Program of China (Grant Nos. 11074263, 10874194, 10734080), the Doctor Fund of (No.B2011-076) and the High-Performance Grid Computing Platform of Henan Polytechnic University.